\documentclass{ws-ijmpcs}
\usepackage{yllmath,color,subfigure}
\providecommand{\omegados}{{\omega_\text{dos}}}

\begin{document}

%
\catchline{}{}{}{}{}
%

\title{Aspects of Localization across the 2D Superconductor-Insulator Transition}
\author{NANDINI TRIVEDI$^1$, YEN LEE LOH$^2$, KARIM BOUADIM$^3$, and MOHIT RANDERIA$^1$}
\address{
	$^1$Department of Physics, The Ohio State University, Columbus, OH  43210, USA\\
	E-mail: trivedi.15@osu.edu\\
	http://www.physics.ohio-state.edu/$\sim$trivedi
}
\address{
	$^2$Department of Physics and Astrophysics, University of North Dakota\\
	Grand Forks, ND  58202, USA
}
\address{$^3$Institute for Theoretical Physics III, University of Stuttgart, Germany}
\maketitle


\begin{abstract}

It is well known that the metal-insulator transition in two dimensions for non-interacting fermions takes place at infinitesimal disorder.  In contrast, the superconductor-to-insulator transition takes place at a finite critical disorder (on the order of $V_c \sim 2t$), where $V$ is the typical width of the distribution of random site energies and $t$ is the hopping scale.  In this article we compare the localization/delocalization properties of one and two particles.  Whereas the metal-insulator transition is a consequence of single-particle Anderson localization, the superconductor-insulator transition (SIT) is due to pair localization -- or, alternatively, fluctuations of the phase conjugate to pair density. The central question we address is how superconductivity emerges from localized single-particle states. We address this question using inhomogeneous mean field theory and quantum Monte Carlo techniques and 
make several testable predictions for local spectroscopic probes across the SIT.
We show that with increasing disorder, the system forms superconducting blobs on the scale of the coherence length embedded in an insulating matrix. In the superconducting state,
the phases on the different blobs are coherent across the system whereas in the insulator long-range phase coherence is disrupted by quantum fluctuations.
As a consequence of this emergent granularity, we show that the single-particle energy gap in the density of states survives across the transition, but coherence peaks exist only in the superconductor.  A characteristic pseudogap persists above the critical disorder and critical temperature, in contrast to conventional theories. Surprisingly, the insulator has a two-particle gap scale that vanishes at the SIT despite a robust single-particle gap.

\end{abstract}


\ccode{PACS numbers:74.25.-q,72.15.Rn,02.70.Ss,64.70.Tg, 05.30.Rt,42.50.Lc}

\section{Introduction}

A superconductor (SC) is an emergent state of matter in which electrons pair up forming Cooper pairs, the different Cooper pairs become phase coherent, and the system undergoes Bose-Einstein condensation. What is the effect of disorder on such a phase-coherent state? It was argued by Anderson~\cite{anderson1959} that three-dimensional superconductivity is quite robust, persisting even in polycrystalline or amorphous materials. Two dimensions turns out to be particularly intriguing because it is the marginal dimension for localization and superconductivity. One can ask the question the other way around: starting with a two-dimensional disordered system in which all the single-particle states are localized, how does it develop superconductivity when attractive interactions between electrons are turned on? What is the specific mechanism~\cite{ma1985} that generates superconductivity in a localized system?
It is seen from experiments that superconductivity in two dimensions does exist but can be destroyed by a large variety of tuning parameters including temperature, inverse thickness (characterized by sheet resistance), disorder, gate voltage, Coulomb blockade, perpendicular magnetic field, and parallel magnetic field~\cite{strongin1970,haviland1989,valles1989,valles1992,hebard1990,shahar1992,chervenak1999,steiner2005,stewart2007,nguyen2009,lee2011,lin2011,bollinger2011}.  The destruction of  superconductivity is a quantum phase transition~\cite{sachdev_qpt}  occurring at zero temperature. 
In this article we provide answers to the following questions related to the superconductor-insulator transition (SIT):

\begin{enumerate}
\item For zero disorder, we know that above the superconducting transition temperature $T_c$ the system is a normal Fermi liquid.  What is the nature of the state above $T_c$ at finite disorder?
Is it a Fermi liquid?

\item What is the nature of the insulator?  Is it a localized Anderson insulator, a Mott insulator, a Fermi glass, a Bose glass, or something else?

\item Is multifractality of the single particle states important at the SIT?

\item What are the energy scale(s) in the insulator that vanish at the SIT?

\item What is the mechanism that drives the SIT?

\item How do the single-particle spectral functions and dynamical conductivity behave in the superconductor, the insulator, and near the SIT?

\end{enumerate}

We start with a definition of the Anderson model of localization and the nature of non-interacting states at different energies. We then discuss the disordered attractive Hubbard model
and the nature of many-particle states obtained by Bogoliubov-de Gennes mean field theory~\cite{ghosal1998,ghosal2000}. We augment the inhomogeneous mean field theory with quantum Monte Carlo simulations and maximum entropy techniques to extract
one-particle and two-particle spectral information~\cite{bouadim2011}. We specifically address the role of amplitude variations in a random environment and
phase fluctuations of the order parameter in generating superconducting and insulating phases. We conclude with ideas for future experiments. 

\section{Anderson Model of Localization} \label{anderson}
Consider a tight-binding model with a disorder potential $v_i$ at each site, chosen independently from a uniform distribution on $[-V, +V]$, where $V$ is the disorder strength.
This is known as the ``Anderson model'' of localization:
	\begin{align}
	H
	&=
	-	\sum_{ij} t_{ij} \cdag_{i} \cccc_{j}
	+	\sum_{i} (v_i - \mu) n_{i}
	.
	\end{align}
The hopping alone would produce plane-wave eigenstates with a bandwidth of $2zt$, where $z$ is the coordination number,
whereas the disorder potential alone would produce site-localized eigenstates with a bandwidth of $2V$.  The competition between hopping and disorder makes this a non-trivial problem.

	\begin{figure}[!ht]
	\centering
	\includegraphics[width=4.5in]{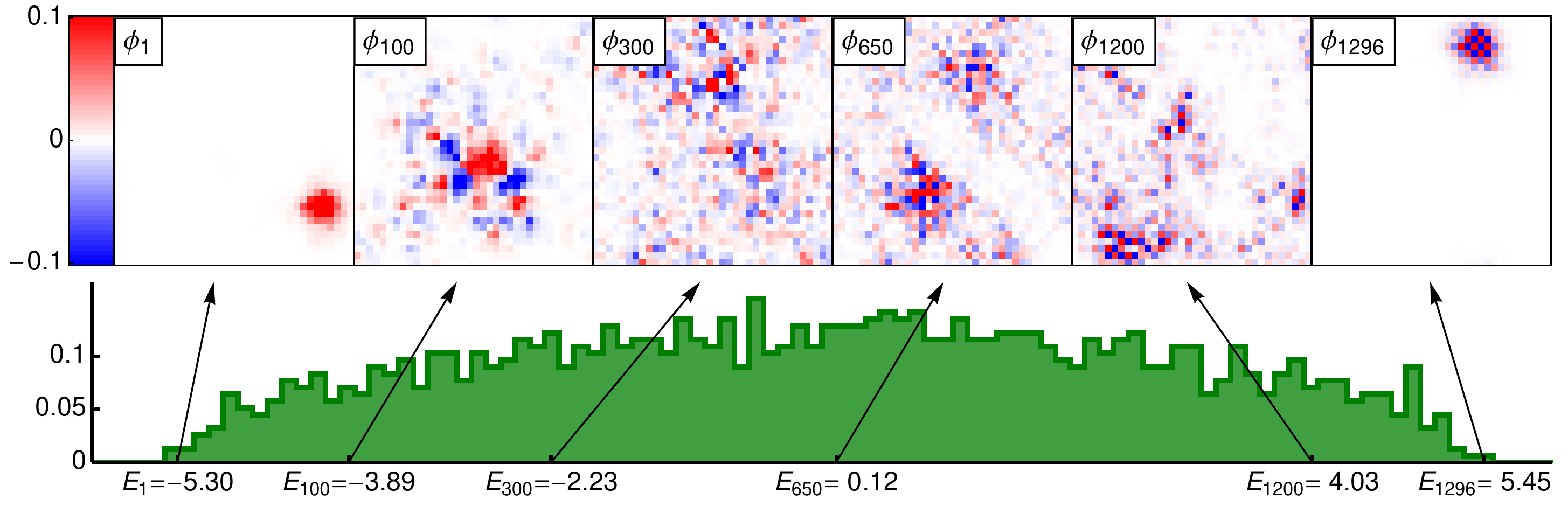}
	\caption{
		Six eigenstates of the Anderson model on a $36\times 36$ square lattice
		for a single disorder realization ($\mu=0, V=3t$).
		Red and blue colors indicate signs of eigenfunctions $\psi_{i\alpha}$.
	}
	\label{anderson-evecs}
	\end{figure}

Figure~\ref{anderson-evecs} shows selected eigenvectors of the Anderson model for a given realization of the random potentials. The eigenvectors correspond to energies in the
tails and in the middle of the density of states. The states in the band tail are localized, whereas the states in the band center appear to be extended over the size of the system; however, in an infinite-size system, they would be localized~\cite{PhysRevLett.42.673,Lee-TVR}.

For a two-dimensional system with on-site disorder, it is well known that an infinitesimal disorder strength, $V$, is enough to localize all single-particle eigenstates.  That is, 
there is a metal-insulator transition occurring at infinitesimal $V$.

\begin{figure}[!ht]\centering
		\includegraphics[width=0.5\textwidth]{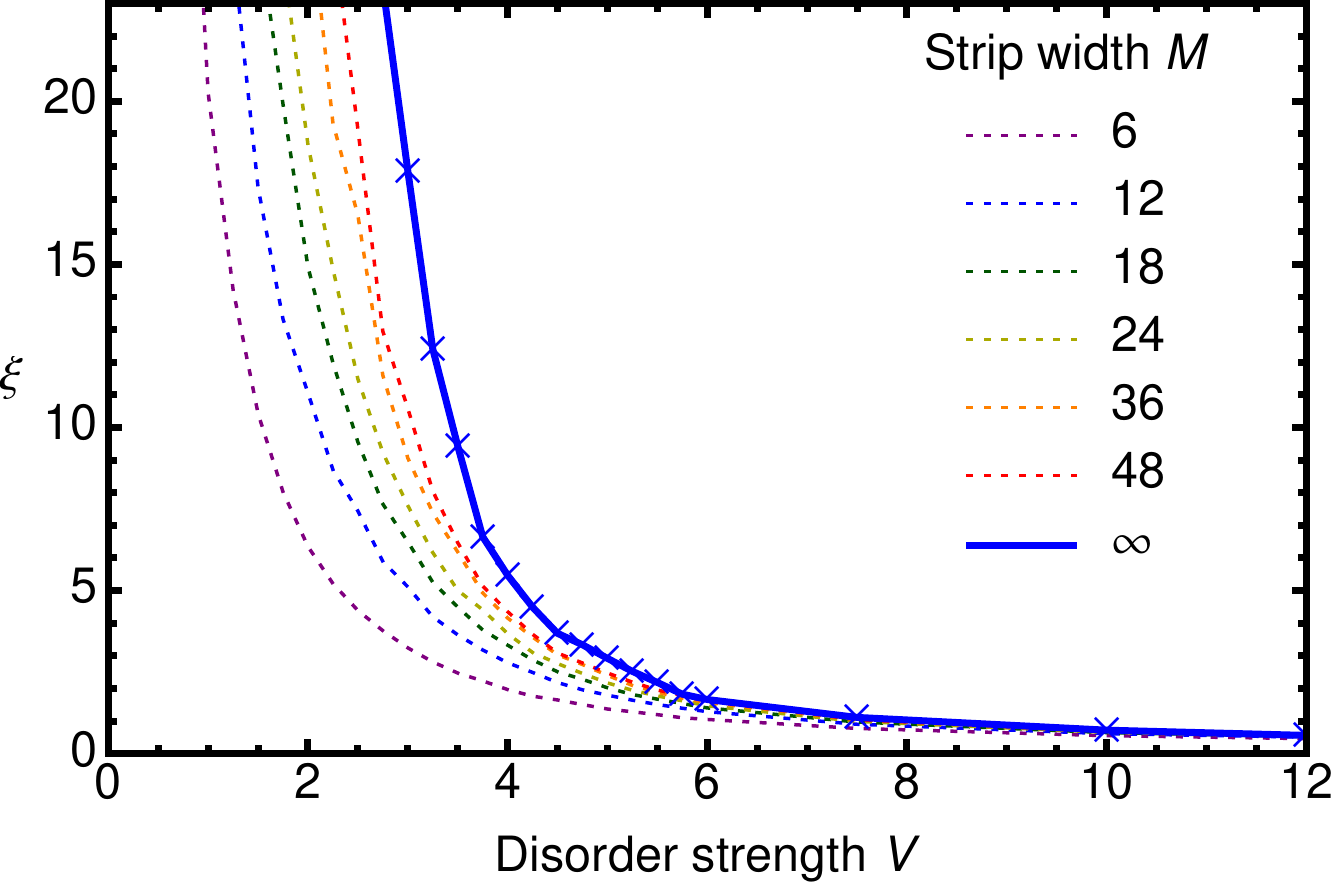}
	\caption{
		Localization length of the 2D Anderson model (blue curve),
		estimated using a crude form of finite-size scaling 
		on Green functions of long strips.
	}
	\label{andersonLocLen.pdf}
	\end{figure}

The localization length $\xi_\text{loc}$ is the length scale for the exponential decay of an eigenfunction far from its center of mass.
It can be obtained from the decay length of the transmission coefficient along a long strip calculated using transfer matrix methods or Green function methods\cite{mackinnon1981}.  
The localization length $\xi_\text{loc}$ is finite for any finite $V$ and decreases as $V$ increases (see Fig.~\ref{andersonLocLen.pdf}).

\section{Disordered Attractive Hubbard Model}
We now turn to the main topic of this paper: a disordered superconductor.  We represent this by an attractive Hubbard model with a disorder potential.  Alternatively, one may think of this as the Anderson model of localization plus an on-site Hubbard attraction, $\left|U\right|$:
	\begin{align}
	H
	&=
	-	\sum_{ij\sigma} t_{ij} \cdag_{i\sigma} \cccc_{j\sigma}
	- \sum_{i} \mu_i n_{i\sigma}
	-	\left|U\right|\sum_{i} \cdag_{i\up} \cdag_{i\dn} \cccc_{i\dn} \cccc_{i\up}
	,
	\end{align}
where	$\mu_i = v_i - \mu$, where the disorder potential at each site $v_i$ is picked independently from a uniform distribution on the interval $[-V,+V]$, as before.

The attractive interaction $\left|U\right|$ has profound consequences.  In the absence of disorder, it is well known that an infinitesimal $\left|U\right|$ is sufficient to produce Bardeen-Cooper-Schrieffer (BCS) pairing, which leads to the phenomenon of superconductivity.
The superconducting state, moreover, exhibits quasi-long-range order up to a finite temperature $T_\text{BKT}$, where it is destroyed by a Berezinskii-Kosterlitz-Thouless (BKT) transition involving vortex-antivortex unbinding.

\subsection{Atomic Limit} 
\begin{figure}
		\centering
		\includegraphics[width=0.45\textwidth]{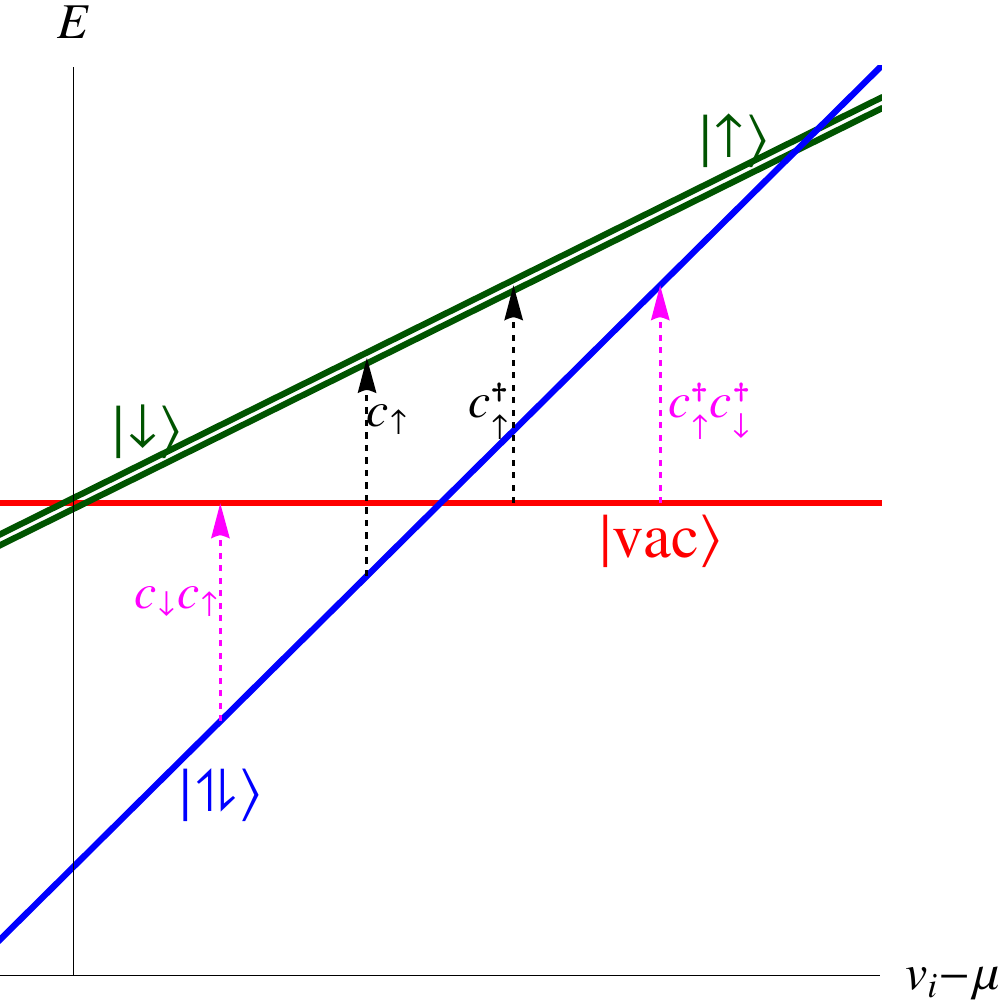}
		\includegraphics[width=0.45\textwidth]{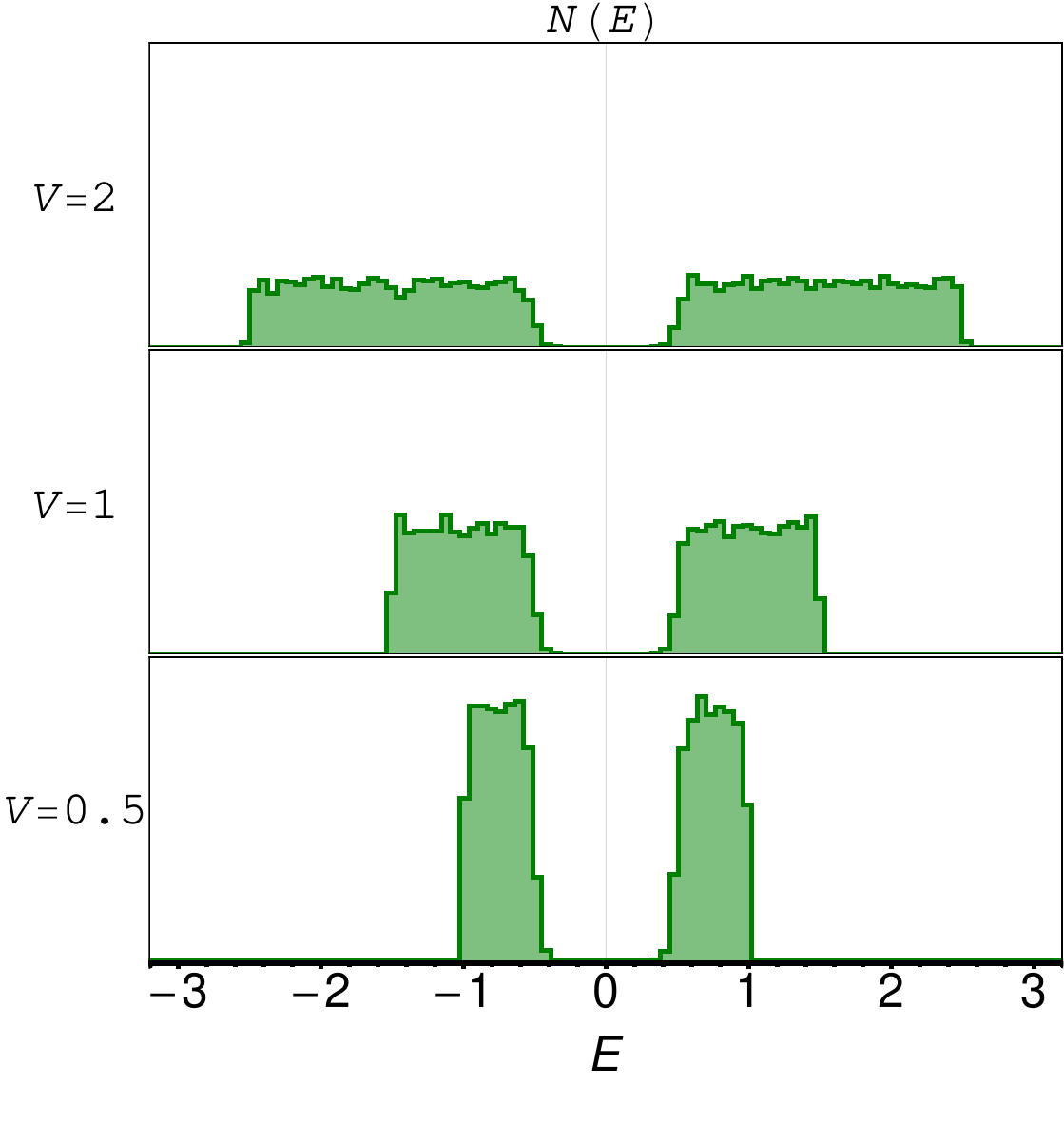}
	\caption{
		\label{AtomicLimit}
		(Left) Energy levels $E$ and single-particle and two-particle transition energies
			for a single site with potential $v_i$, chemical potential $\mu$,
			and attraction $\left|U\right|$ -- the atomic limit of the Hubbard model.
		(Right) Density of states of a $100\times 100$ lattice in the atomic limit ($t=0$)
		 with attraction $\left|U\right|=1$.  
		 At zero temperature the DoS has a hard gap $\left|U\right|/2$ regardless of disorder strength $V$.
		 At finite temperature ($T=0.05$ in the figure above) there is a small amount of weight in the gap due to thermal excitations.
	}
	\end{figure}

\slabel{atomicLimit}
In the limit of extreme disorder, the hopping can be neglected, and the system then reduces to an ensemble of single-site Hubbard models, each with the Hamiltonian
	\begin{align}
	H
	&=-\left|U\right| n_{\up} n_{\dn}	+ (V - \mu) ( n_{\up} + n_{\dn}	)
	.
	\end{align}
This system has just four Fock states.  The energies of these states are
	$	E_{0}	= 0	$,
	$ E_\up = E_\dn = V - \mu $, 
	and
	$	E_{\up\dn} = -\left|U\right| + 2(V - \mu) $
	.
The four states occur with relative Boltzmann weights $\exp (-\beta E_n)$.
The spectral function (the density of states for single-particle excitations) can be obtained by considering transitions between these four Fock states (amplitudes and energies).
This is illustrated in Fig.~\ref{AtomicLimit}.
The ground state is always either doubly occupied or empty.
Regardless of the on-site potential $V$, single-particle transitions (black arrows) always cost at least $|U|/2$, and therefore the spectrum is always gapped.
\footnote{This is in contrast to the repulsive-$U$ Hubbard model in the atomic limit, for which the spectrum is only gapped if $U > 2V$.
}
Pair excitations (purple arrows) correspond to transitions from the doubly occupied state to vacuum or vice versa.
At the specific matching value $\mu_i = v_i - \mu=0$ these pair excitations may cost zero energy.
We have generalized the above calculation to exact diagonalization of the many-body Hubbard Hamiltonian on small clusters of a few sites, which leads to the same conclusions: single-particle excitations are gapped whereas two-particle excitations can become gapless for an appropriate choice of $\{v_i - \mu\}$.

We know that a clean s-wave superconductor ($V=0$) has a gap $E_g = \Delta$ given by the BCS gap equation.
We have just found that in the limit of extreme disorder, $V \gg (\left|U\right|,t)$, the gap is finite and large,  $E_g = \left|U\right|/2$.
We willl see that the gap remains finite between these two extremes.

\subsection{Pairing of Exact Eigenstates (PoEE)}

The above Hamiltonian contains three terms: hopping, disorder, and attraction.  
In typical $s$-wave superconductors, the first two terms have the largest energy scales.
Thus, it makes sense to solve the non-interacting problem first by direct diagonalization, to find the 
disorder eigenvalues and eigenstates $\xi_\alpha$ and $\phi_{i\alpha}$ and then examine the effect of $\left|U\right|$. 
This is very much in the spirit of the derivation of Anderson's theorem~\cite{anderson1959}.
	 
In the basis of exact eigenstates, the Hamiltonian is 
	\begin{align}
	H
	&=
	 \sum_{\alpha}	\xi_\alpha \gamma^\dag_{\alpha\sigma} \gamma_{\alpha\sigma}
	-	\left|U\right|\sum_{\alpha\beta\gamma\delta i}			
		\phi_{i\alpha} \phi_{i\beta} \phi^*_{i\gamma} \phi^*_{i\delta}
		 \cdag_{\alpha\up} \cdag_{\beta\dn} \cccc_{\gamma\dn} \cccc_{\delta\up}
	.
	\end{align}
Following Anderson's suggestion, let us assume that instead of pairing between $\kkk$ and $-\kkk$, we have pairing between time-reversed eigenstates $\alpha$ and $\bar\alpha$ (i.e., complex conjugate eigenfunctions).
Retaining only those terms in the Hamiltonian that connect such eigenstates we obtain the gap equation~\cite{ghosal1998,ghosal2000}
	\begin{align}
	H_\text{PoEE}
	&=
	 \sum_{\alpha}	\xi_\alpha \beta^\dag_{\alpha\sigma} \beta_{\alpha\sigma}
	-	\left|U\right|\sum_{\alpha\beta i}			
		\phi_{i\alpha} \phi_{i\bar\alpha} \phi^*_{i\bar\beta} \phi^*_{i\beta}
	 \cdag_{\alpha\up} \cdag_{\bar\alpha\dn} \cccc_{\bar\beta\dn} \cccc_{\beta\up}
	\nonumber\\
	&=
	 \sum_{\alpha}	\xi_\alpha \beta^\dag_{\alpha\sigma} \beta_{\alpha\sigma}
	-	\sum_{\alpha\beta}			
		M_{\alpha\beta}
	 \cdag_{\alpha\up} \cdag_{\bar\alpha\dn} \cccc_{\bar\beta\dn} \cccc_{\beta\up}
	\end{align}
where
$		M_{\alpha\beta}
		=
\left|U\right| \sum_i \left| \phi_{i\alpha} \right|^2  \left| \phi_{i\beta} \right|^2
$
.
Approximate this by a mean-field Hamiltonian 
	\begin{align}
	H_\text{MF}
	&=
	 \sum_{\alpha}	\xi_\alpha \beta^\dag_{\alpha\sigma} \beta_{\alpha\sigma}
	-	\sum_{\beta} 
	(
		\Delta^*_\beta
    \cccc_{\bar\beta\dn} \cccc_{\beta\up}
  +h.c.
  )
	\end{align}
(up to a constant), where the order parameter is
	\begin{align}
	\Delta^*_\beta
	&=
		\left|U\right|
		\sum_{\alpha}
		M_{\alpha\beta}
	\mean{	 \cdag_{\alpha\up} \cdag_{\bar\alpha\dn}   }
	\end{align}
(assuming that $\xi_\alpha$ have been redefined in this step to include Hartree shifts).

The gap equation works out to be
	\begin{align}
	\Delta_\alpha
	&=
		\left|U\right|\sum_{\beta}
		M_{\alpha\beta}
	\frac{\Delta_\beta}{2E_\beta}
	\tanh \frac{E_\beta}{2T}
	\end{align}
where
$
	E_\beta
	=	\sqrt{\xi_\beta{}^2 + \Delta_\beta{}^2} 
$,
and the chemical potential is determined by the number equation
	\begin{align}
	\mean{n}
	&=	\frac{1}{N}
		\sum_\alpha \left( 1 - \frac{\xi_\alpha}{E_\alpha} \right)
		.
	\end{align}

The PoEE theory can be used in the above form, or one can perform further approximations as follows.  In the low-disorder regime, the disorder eigenstates $\phi_{i\alpha}$ are extended on the scale of the system, so that $M_{\alpha\beta} \approx 1/N$ independent of $\alpha$ and $\beta$.  In this limit Anderson's theorem applies -- the gap equation takes the simple BCS form, and $\Delta$ is spatially uniform.  In the high-disorder regime, on the other hand, the disorder eigenstates are strongly localized with localization lengths $ \xi^\text{loc}_\alpha$, and the $M$ matrix is approximately diagonal,
$M_{\alpha\beta} \approx \delta_{\alpha\beta} \sum_i \left| \phi_{i\alpha} \right|^4
\approx 
\delta_{\alpha\beta} / (\xi^\text{loc}_\alpha)^2
$.

Surprisingly, \emph{the gap is finite for all values of disorder} $0<V<\infty$.  At large disorder, the gap increases with disorder as
	\begin{align}
	E_g
	&=	\frac{\left|U\right|}{2\xi_\text{loc} {}^2}
	,
	\label{gapPoEE}
	\end{align}
where $\xi_\text{loc}$ is the localization length at the chemical potential: as the single-particle states become more localized, the effective attraction is enhanced, leading to a larger gap.  
\footnote{The model Hamiltonian does not include Coulomb repulsion.  In real materials it is possible that the Finkel'stein mechanism -- disorder-enhanced Coulomb repulsion -- may compete with the disorder-enhanced attraction mechanism described here.  A full analysis remains to be done.
}

There have been proposals~\cite{feigelman2007,fractalannals} that the multifractal nature of the single particle states modifies the exponent in Eq.~\eqref{gapPoEE} to the fractal dimension $d_f=1.7$ in 3D. It is not entirely clear how applicable the fractal nature of the eigenstates is 
in two dimensions, where the metal-insulator transition which occurs at $V^\text{MIT}_c=0^+$ and the superconductor-insulator transition at $V^\text{SIT}_c\approx 2t$ are widely separated.  More importantly, it should be remembered that this is a mean-field analysis and there are considerably more important changes to the many-particle states introduced by phase fluctuations. 

It has also been proposed~\cite{mirlin,fractalannals} that $T_c$ gets enhanced near the SIT and can in fact increase without bound. These statements and calculations assume that the SC transition temperature is determined by the gap which as we have discussed is incorrect.
Within BCS theory in a weakly coupled clean SC the transition temperature $T_c$ is indeed determined by the gap scale. But that situation changes entirely in a strongly coupled SC or even in a weakly coupled but disordered SC as we have shown. The gap remains finite across the SIT but the phase stiffness 
goes soft and ultimately vanishes at the transition and it is the phase stiffness scale that now determines $T_c$. So while a temperature scale associated with the gap may increase near the SIT, the true transition $T_c$ at which the resistance vanishes decreases monotonically with disorder
and vanishes at the SIT.

The PoEE approach is useful for understanding the robustness of the gap,
but it does not give the full story.
It predicts the BCS coherence peaks in the density of states survive up to infinite disorder, whereas more accurate calculations show that there are significant pile-ups in the density of states only in the superconductor. 
Furthermore, being a mean-field theory, PoEE fails to capture the destruction of phase coherence at SIT due to quantum phase fluctuations.  We now proceed to more detailed discussion of phase fluctuations.

	\begin{figure}[!ht] \centering
	\includegraphics[width=4.5in]{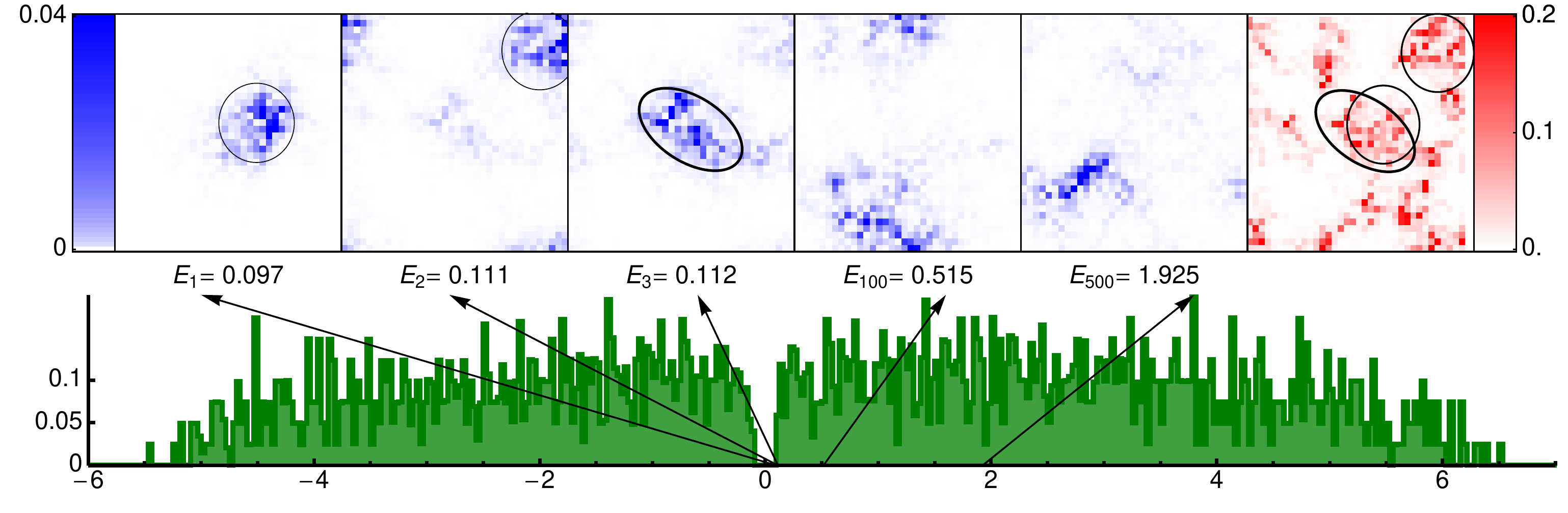}
	\caption{
		Eigenvalues and eigenvectors of a disordered superconductor
		(see text for description of model).
		The first five panels show the magnitude of five BdG eigenstates (bogolon wavefunctions),
			$\left|u_i\right|^2 + \left|v_i\right|^2$.
		The last panel (red) is a map of the local pairing amplitude $\Delta_i$.
		The low- and high-energy eigenstates are localized, 
		whereas the intermediate-energy eigenstates are quasi-extended.
		In particular, the lowest eigenstates correspond to
		 the locations of the superconducting puddles (where $\Delta_i$ is large).
		The parameters are $\left|U\right|=1.5t$, $n=0.875$, $N=36\times 36$, $V=3t$.		
	}
	\label{bdg-evecs}
	\end{figure}

\section {BdG and QMC results:}

\subsection {Eigenstates:}

Let us consider the eigenvalues and eigenvectors of a dirty superconductor within Bogoliubov-de Gennes (BdG) inhomogeneous mean-field theory, as shown in Fig.~\ref{bdg-evecs}.
The eigenvectors are considerably more localized than for the non-interacting case (Fig.~\ref{anderson-evecs}).
This is clearly seen when comparing Fig.~\ref{anderson-evecs} $(U=0)$ and Fig.~\ref{bdg-evecs} $(\left|U\right|=1.5t)$, which are both at the same disorder strength, $V=3t$.
The eigenvalue spectrum is gapped. How then does this system sustain superconductivity?

We will see that both amplitude and phase fluctuations play an integral role in driving the SIT.

	\begin{figure*}[!ht] \centering
	\includegraphics[width=0.8\textwidth,clip,trim=0 0 90mm 0]{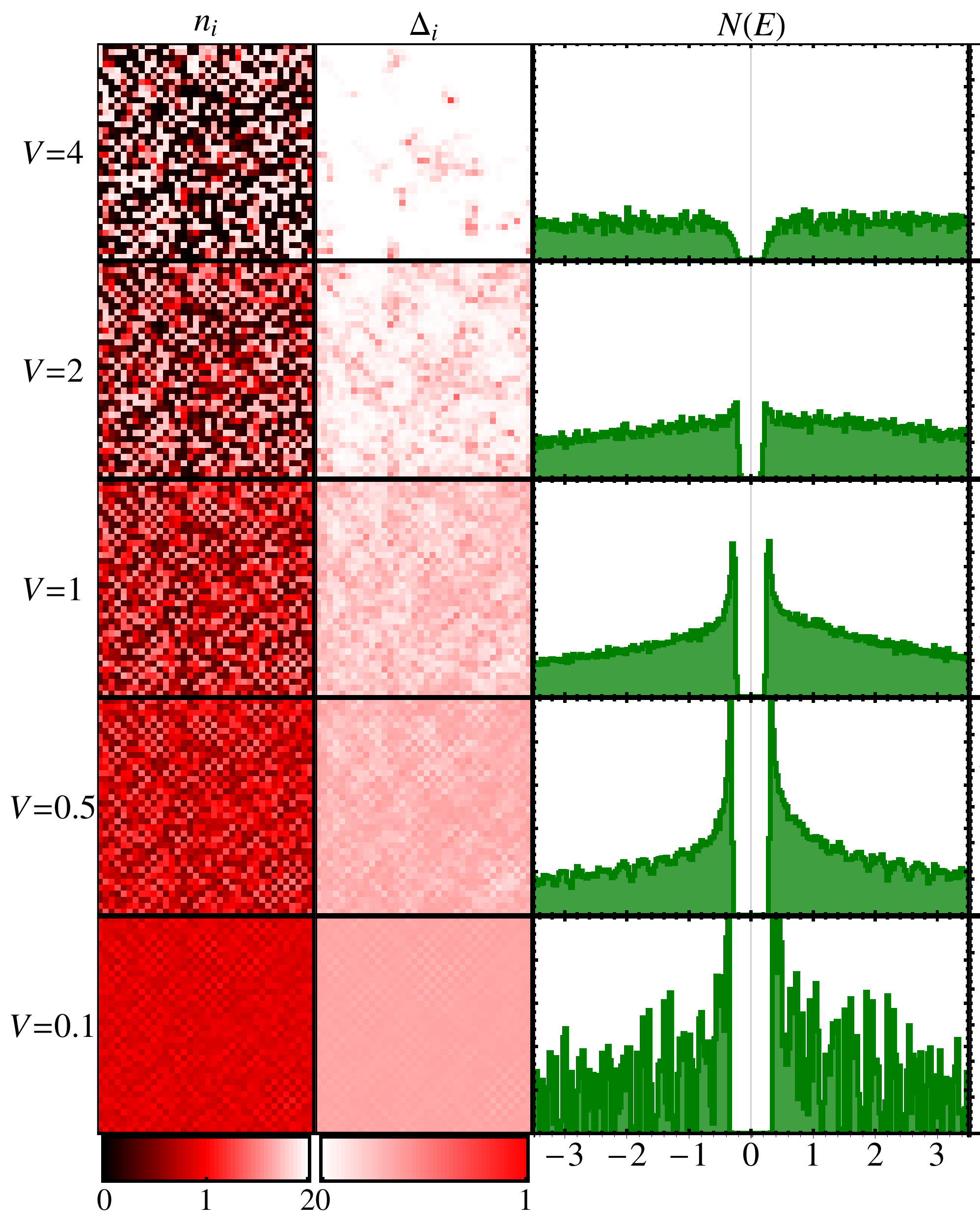}
	\caption{
		The three columns show BdG results for the number density, pairing density, and
			single-particle spectrum
			of the attractive Hubbard model with a disorder potential,
			on a $36\times 36$ square lattice 
			with $\left|U\right|=2$, $\mean{n}=0.875$, and $T=0$,
			as functions of disorder strength $V$.  Here $t=1$.
	}
	\label{ghosal-everything}
	\end{figure*}

	\begin{figure}[!ht]	\centering
		\includegraphics[width=16pc]{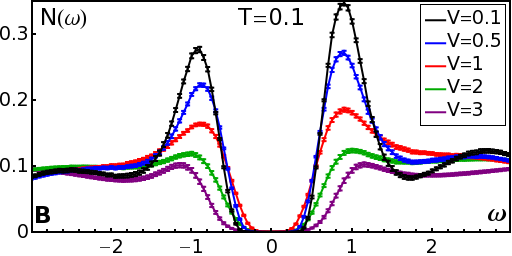}
		\caption{
		\label{DisorderDependenceQMCMEM} 
		Disorder dependence of single-particle spectrum from QMC+MEM
		at very low temperature.
		With increasing disorder, 
		quantum phase fluctuations eventually wash out the coherence peaks,
		but the gap is robust.
		}		
	\end{figure}

\subsection {Amplitude {\it vs.} Phase Fluctuations:}
A singlet $s$-wave superconductor is described by a complex order parameter 
$\Delta(\RRR) = \left| \Delta(\RRR) \right|  e^{i\theta(\RRR)}$.
At zero temperature the pairing amplitude $\left| \Delta(\RRR) \right|$ takes a uniform value, $\Delta_0$.
This is the energy scale associated with pairing.
It typically manifests itself as an energy gap $E_{g}=\Delta_0$,
and it also sets the maximum temperature, $T_\text{pair} = 0.57 \Delta_0$, 
for the formation of Cooper pairs.

On the other hand, the fluctuations of the phase, $\theta(\RRR)$, are controlled by the superfluid density (or phase stiffness) $\rho_s$.  For 2D superconductors $\rho_s$ has the dimensions of energy, and it can be directly interpreted as the energy scale for phase fluctuations.
$\rho_s$ can be measured using mutual inductance techniques.
It also sets the maximum temperature, $T_\text{phase}$, for long-range phase coherence.

What will emerge from the discussions below is that for a clean superconductor $(V\rightarrow 0)$, the single-particle eigenstates are ``localized'' by attraction, on the scale of the superconducting \emph{coherence length} $\xi_\text{coh}^0$.  
At weak disorder the pairing amplitude is homogeneous.  At strong disorder the system breaks up into blobs as seen in Fig.~\ref{bdg-evecs} and the
system can be described by a granular superconductor or a Josephson Junction Array (JJA), where phase fluctuations are extremely important.

What is the size of these blobs? It is given by the coherence length $\xi_\text{coh}$ defined as the scale over which the order parameter bends and is modified to include the effects of disorder.
In the limit of large disorder $(V\rightarrow \infty)$, the single-particle eigenstates are dominated by Anderson localization, and are localized on the scale of $\xi_\text{loc}$.  Thus $\xi_\text{blob}={\rm min} [\xi_\text{coh},\xi_\text{loc}]$.
For weak disorder, the phases of the order parameter on different blobs are coupled leading to a globally phase coherent superconducting state. 
On the other hand at strong disorder the phases on the different blobs lose long range phase coherence and the system becomes an insulator (see Fig.~\ref{GhosalPuddleCartoon}).
As the quantum phase transition is approached both the correlation length $\xi \approx  \xi_\text{blob}\delta^\nu$ and the correlation time $\xi_\tau \approx  \xi_\text{blob}\delta^{z \nu}$ diverge. Here $\nu$ and $z$ are critical exponents.

\begin{figure}[htb]
\centering
\begin{minipage}{20pc}
\includegraphics[width=20pc,clip,trim=0 6pc 0 6pc]{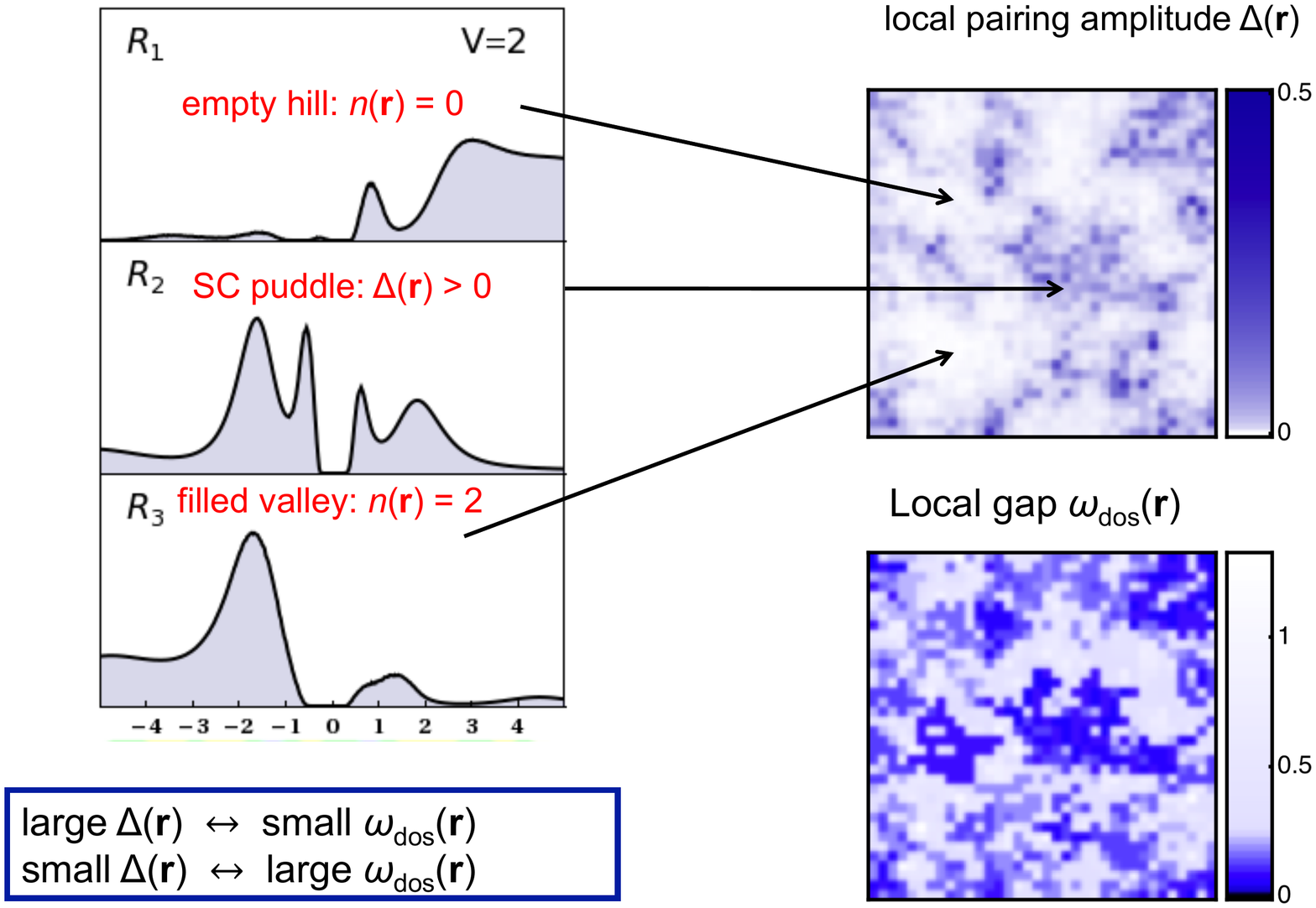}
\caption{\label{LocalSpecAndLocalGap} 
	(Right) Within BdG, the local pairing amplitude is \emph{anticorrelated} with the local gap.
	(Left) LDOS results from QMC+MEM.
	Site ${\bf R_1}$ is on a high potential hill that is nearly empty, 
	and ${\bf R_3}$ is in a deep valley that is almost doubly occupied. 
	This leads to the characteristic asymmetries 
	in the LDOS for ${\bf R_1}$ and ${\bf R_3}$. The small
	local pairing amplitude $\Delta(\RRR)$ at these two sites is reflected in the absence of coherence peaks in the LDOS.
	In contrast, site ${\bf R_2}$ has a density closer to half-filling,
	leading to a significant local pairing amplitude, a much more symmetrical LDOS, 
	and coherence peaks that persist even at strong disorder.
}
\end{minipage}

\bigskip

\bigskip

\hspace{1pc}
\begin{minipage}{14pc}
\includegraphics[width=14pc]{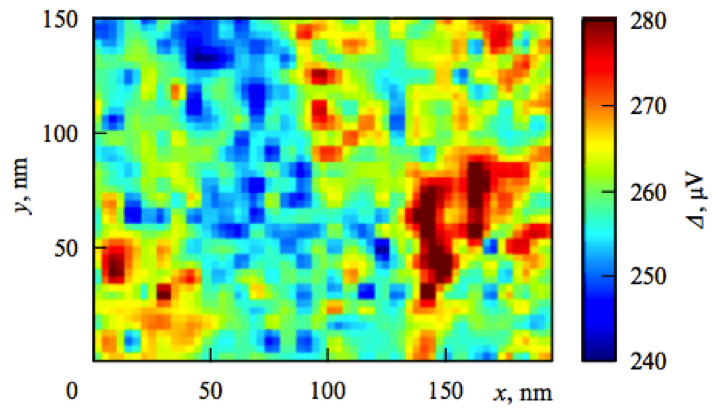}
\caption{Gap map of a TiN film obtained from scanning tunneling spectroscopy (from Sacepe et al., 2011) ${}^{27}$, showing inhomogeneities on a scale of a few tens of nanometers.}
\label{SacepeGapMap} 
\end{minipage}
\end{figure}

\subsection {Emergent Granularity:}

In the Finkel'stein mechanism~\cite{finkelstein1994} of the superconductor-insulator transition, the pairing amplitude and the gap decrease with increasing disorder due to enhancement of Coulomb repulsion, and both fall to zero at the SIT.  However, Finkel'stein's analysis assumes that the pairing amplitude is uniform across the system.  

BdG calculations show that as $V$ approaches $V_c$, the pairing amplitude $\Delta(\rrr)$ becomes strongly inhomogeneous, as shown in Fig.~\ref{ghosal-everything}.  The superconductor breaks up into puddles, on the scale of the coherence length, where $\Delta(\rrr)$ is large; in the surrounding regions, $\Delta(\rrr)$ is negligible.  Such puddles still exist even at large $V$.

\subsection {Pairing amplitude versus single-particle gap:}

Meanwhile, the single-particle gap $E_\text{gap}$ remains finite (see Figs.~\ref{ghosal-everything} and \ref{DisorderDependenceQMCMEM}).  This is \emph{not} an artifact of BdG; it is a very robust conclusion that is confirmed by quantum Monte Carlo (QMC) combined with maximum entropy methods (MEM) to extract spectral behavior. 
\footnote{One might think that Griffiths-McCoy-Wu rare regions might produce subgap weight.  We have found that this effect is insignificant, and in any case, in two dimensions they do not affect the conclusion of a robust single-particle gap.  Rare region effects will be dealt with in a forthcoming paper.}

Figure~\ref{LocalSpecAndLocalGap} shows that there is in fact an {anti-correlation} between the local pairing amplitude and the local spectral gap. The SC puddles on which the local $\Delta(\RRR)$ is finite have a finite 
gap with symmetric line shapes and sharp coherence peaks or pile-ups in the density of states at the gap edges. On the other hand, the insulating regions have $\Delta(\RRR)\approx 0$ and very asymmetric broad density of states showing a much larger gap. Although the local gap extracted from the local density of states (LDOS) 
is highly inhomogeneous, it is nevertheless finite at every site,
similar to the experimental data in Fig.~\ref{SacepeGapMap}.

The DOS is the LDOS averaged over all sites.  The gap in the DOS, $E_{gap}$, is the lowest gap in the LDOS on any site.
According to BdG (Fig.~\ref{ghosal-everything}) and QMC (Figs.~\ref{DisorderDependenceQMCMEM}, \ref{PhaseDiagrams}) calculations, $E_{gap}$ remains robust across the SIT, even when thermal and quantum phase fluctuations are included.  Thus the SIT is a transition from a gapped superconductor to a gapped insulator.

The key reason for the robustness of the spectral gap even for high disorder is because of the disorder-induced ``emergent granularity''. The formation of blobs with finite local pairing amplitude leads to regions with a finite local gap.
Further the spectra on these locally SC blobs are fairly symmetric because of strong number fluctuations and particle hole mixing.
These regions are separated from insulating seas where the pairing amplitude is almost zero but not the local gap. The insulating regions arise form either deep valleys that are filled or high hills that are empty
as seen in Fig.~\ref{LocalSpecAndLocalGap}. Given the almost fixed number of particles (either two or zero) these regions allow for very small number fluctuations and hence considerably enhanced phase fluctuations of the conjugate variable
leading to an almost zero pairing amplitude. The gap in these regions is the energy difference between the chemical potential and the local energy in the valley or the hill.  This gap is considerably larger than that on the 
SC blobs and shows an asymmetric line shape because of the lack of particle-hole mixing.

We thus see that all approaches (atomic limit, pairing of exact eigenstates, BdG, and QMC) concur on the existence of a robust, finite gap in the single-particle density of states.

\subsection{Pseudogap over wide temperature range:}
\begin{figure}[h] 	\centering
	\includegraphics[width=0.47\textwidth]{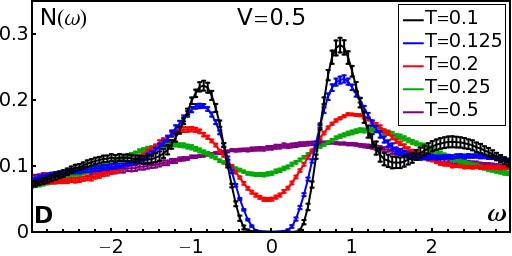}
	\includegraphics[width=0.47\textwidth]{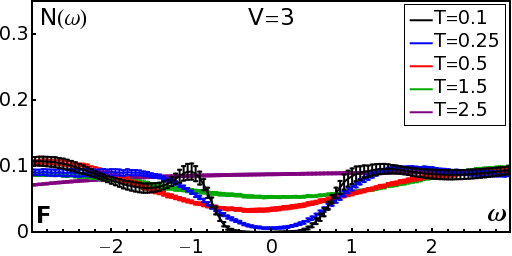}
	\caption{
	\label{TemperatureDependenceQMCMEM} 
	Temperature dependence of DOS from QMC+MEM calculations.
	(Top) At weak disorder, as a function of increasing temperature,
	thermal fluctuations destroy the coherence peaks for $T \gtrsim T_c \approx 0.14$.
	However, a pseudogap remains up to higher temperatures $T \sim 0.4$.
	(Bottom) At strong disorder, there are no coherence peaks;
	there is a hard gap at $T=0$ and a pseudogap up to $T \sim 1.5$.
	}
	\end{figure}

The hard gap at $T=0$ evolves into a pseudogap -- a suppression in the low-energy DOS -- 
which persists well above the superconducting $T_c$ up 
to a crossover temperature scale $T^\ast$, in marked deviation from BCS theory.
This disorder-driven pseudogap also exists at finite temperatures in the insulating state
and grows with disorder (Fig.~\ref{TemperatureDependenceQMCMEM}).
These predictions are in good agreement with experiments~\cite{sacepe2011,mondal2011}.

	\begin{figure}[!ht] 	\centering
	\includegraphics[width=16pc]{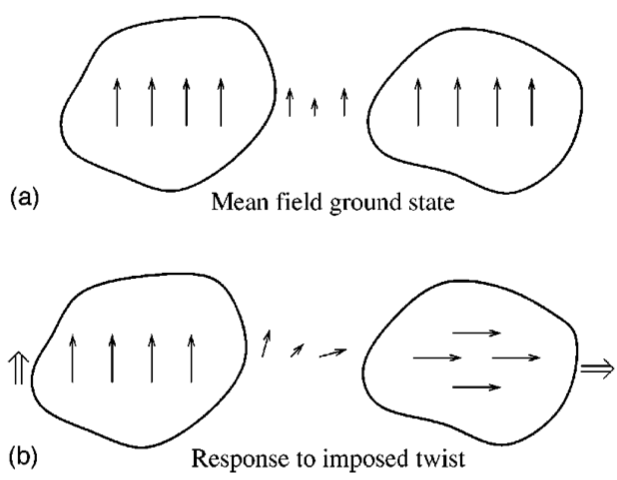}
	\caption{\label{GhosalPuddleCartoon} The superfluid density is a measure of the rigidity of the phases. This rigidity is clearly reduced by thermal fluctuations. It is also reduced by amplitude 
	variations and by quantum phase fluctuations even at $T=0$. The upper panel shows a phase coherent ground state albeit with variations in the amplitude-- large values in the SC puddles and small values in the intervening sea.
	For an applied twist, the system can accommodate most of the twist in regions where the amplitude is small leading to a very small cost in energy and hence a very small superfluid density. }
	\end{figure}
	
\section{Indicators of the phase-fluctuation-driven SIT: superfluid stiffness and off-diagonal long-range order}

If the pairing amplitude and gap are both finite at all $V$, and the single-particle states are all localized, then what is the mechanism of the SIT?

Our BdG and QMC calculations~\cite{bouadim2011} imply that the SIT is driven by \emph{phase fluctuations}.
For low disorder the local pairing amplitude $\Delta({\bf R})\equiv \langle c_{{\bf R}\downarrow}c_{{\bf R}\uparrow}\rangle$ is homogeneous across the system.
However, as disorder is increased the system self-organizes into superconducting blobs on the scale of the coherence length within an insulating matrix
(as seen in Fig.~\ref{ghosal-everything}). The phases of the different blobs are coupled by Josephson tunneling of pairs.
In the globally superconducting state, the phases of the different blobs get locked together whereas in the insulator the phase coherence of the different blobs is lost on ever shorter length and time scales as one moves away from the quantum phase transition. 
This is illustrated schematically in Fig.~\ref{GhosalPuddleCartoon}.

\subsection{Superfluid stiffness}
Ultimately, the SIT is defined by off-diagonal long-range order (ODLRO).  ODLRO manifests itself in the two-particle correlator -- i.e., the amplitude of inserting a pair and removing it at a different time and place.  In other words, ODLRO means that pairs are delocalized and phase coherent, and its absence means that pairs are localized and incoherent.

We have not actually calculated the ODLRO correlator in BdG or QMC.  Rather, we focused on an experimentally measurable quantity, the superfluid stiffness $\rho_s \propto D_s$.  This is determined by the current-current correlator, and it is also a reliable indicator of superconducting long-range order.

\begin{figure}[!ht] 	\centering
	\includegraphics[width=7.5pc]{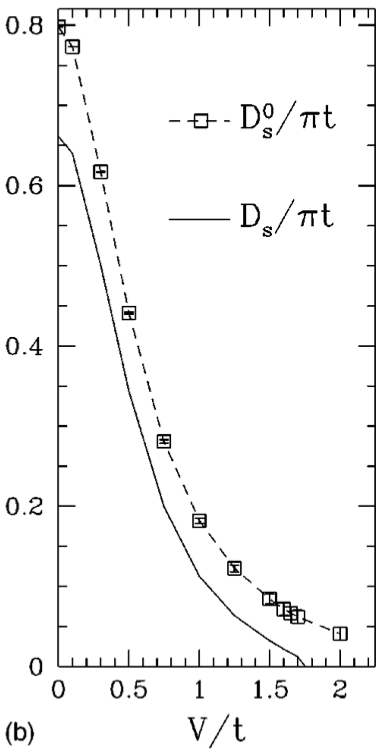}
	\includegraphics[width=14pc]{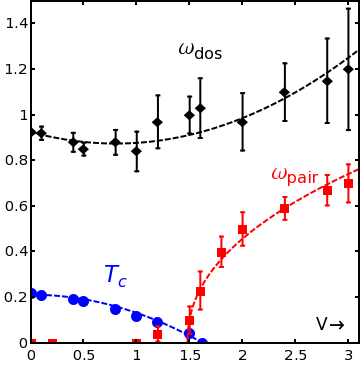}
	\caption{
	\label{PhaseDiagrams}
	(Left)
	Superfluid stiffness according to BdG+SCHA.
	(Right)
	Energy and temperature scales across the superconductor-insulator transition (SIT)
	according to QMC+MEM calculations.
	The single-particle gap $\omegados$ remains finite for all values of disorder $V$,
	whereas the superconducting $T_c$ 
	and the two-particle energy scale $\omega_{\rm pair}$ in the insulator
	both vanish at the SIT.
	}
	\end{figure}

The left panel of Fig.~\ref{PhaseDiagrams} shows the superfluid stiffness according to BdG (dashed curve) and as renormalized within the self-consistent harmonic approximation (SCHA, solid curve).  $D_s$ falls to zero at $V=V_c \sim 1.6t$ (for the given parameters).
The right panel shows results from QMC.  The superfluid stiffness is finite for $V<V_c$, and zero for $V>V_c$; in fact, the QMC simulations give superconductivity in a region of the phase diagram, $T<T_c(V)$.

\subsection{Pair susceptibility}
As seen form Fig.~\ref{PhaseDiagrams}, the transition temperature $T_c$ is suppressed to zero at the quantum phase transition $V=V_c$. What is the energy scale on the insulating side that vanishes at the transition?
We discover the answer in the properties of the two-particle transport $P(\rrr,\rrr',\tau)$.
The local two-particle spectral function, or pair susceptibility $P(\omega)$, is defined as the analytic continuation of the correlation function $P(\tau) = \sum_\RRR \mean{\calT_\tau F (\RRR;\tau) F^\dag (\RRR;0)}$ where $F(\RRR,\tau)=  \cccc_{\RRR\dn} (\tau) \cccc_{\RRR\up} (\tau)$.  Physically, $P(\RRR,\omega)$ is the amplitude for inserting a pair at a site $\RRR$ at energy $\omega$, and $P(\omega)$ is the average insertion amplitude over all sites.

\begin{figure}[!ht] \centering
	\includegraphics[width=10pc]{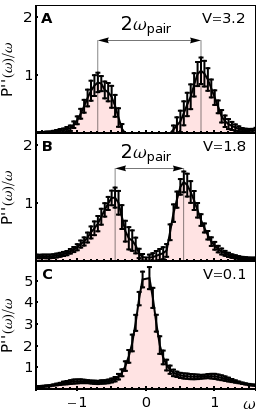}
	\caption{\label{KarimPairSuscep}
	Imaginary part of the dynamical pair susceptibility $P''(\omega) / \omega$ at $T=0.1t$.
	Error bars represent variations between 10 disorder realizations.
	For $V<V_c$ the large peak at $\omega=0$
	 indicates zero energy cost to insert a pair into the SC.
	For $V>V_c$, there is a gap-like structure at $\pm \omega_{\rm pair}$,
	the typical energy required to insert a pair into the insulator.
	}
	\end{figure}

Fig.~\ref{KarimPairSuscep} shows QMC+MEM results for the imaginary part of $P$.
On the superconducting side of the transition there is a large amplitude for inserting pairs at zero energy.
However, on the insulating side, there is a characteristic energy scale $\omega_{\rm pair}$ to insert a pair in the insulator that collapses upon approaching the SIT
(notwithstanding a small amount of spectral weight at low energies coming from rare regions).

Thus the the local two-particle energy scale, $\omega_\text{pair}$, is an indicator of the approach to the SIT on the \emph{insulating} side.  A finite value of $\omega_\text{pair}$ tells us 
about the time scales ($\approx \omega_\text{pair}^{-1}$) over which the pairs are able to travel coherently through the system.  As the transition is approached from the insulating side these time scales get longer
and ultimately diverge at the transition.

\section{Conclusions}

In two dimensions, the metal-insulator transition for the underlying non-interacting disordered problem occurs at $V=0^+$, that is all single-particle states are localized at any finite $V$.
In contrast, the SIT for infinitesimal $\left|U\right|$ occurs at finite $V$.  Thus even with all single-particle states localized, ``long-distance two-particle transport is possible'' 
because of the presence of the condensate in which pairs are delocalized and coherent''.
Our main results are summarized in Fig.~\ref{PhaseDiagrams}.
The superconducting phase persists up to a finite temperature $T_c(V)$, shown in Fig.~\ref{PhaseDiagrams}. Beyond that a pseudogap persists in the density of states though coherence peaks are suppressed at $T_c$.
In this regions there are signatures of pairing but thermal and quantum fluctuations destroy phase coherence.
The insulator is a novel kind of insulator possessing a finite gap to single-particle excitations, which indicates that it consists of bound pairs.  The size of these pairs should be viewed to be on the size of the local SC blobs.
However, in spite of the presence of these pairs or blobs the system is not a SC at a global level because the phases on the different blobs are incoherent due primarily to quantum phase fluctuations
in the proximity of the quantum phase transition.

With these calculations we have nailed the nature of the phases and the mechanism of the disorder-driven quantum phase transition. The next level of open questions are now related to the behavior of the 
frequency-dependent response functions and transport as a function of temperature and disorder across the SIT. 

\subsection{Acknowledgments}
We gratefully acknowledge support from  US Department of Energy, 
Office of Basic Energy Sciences grant DOE DE-FG02-07ER46423 (N.T., Y.L.L.), 
NSF DMR-0907275 (K.B.), NSF DMR-1006532 (M.R.), and computational support from 
the Ohio Supercomputing Center.

\bibliographystyle{ws-procs9x6}


\begin{thebibliography}{10}

\bibitem{anderson1959}
P.~W. Anderson, {\em Journal of Physics and Chemistry of Solids} {\bf 11},
  26(September 1959).

\bibitem{ma1985}
M.~Ma and P.~A. Lee, {\em Phys. Rev. B} {\bf 32}, 5658(Nov 1985).

\bibitem{strongin1970}
M.~Strongin, R.~S. Thompson, O.~F. Kammerer and J.~E. Crow, {\em Phys. Rev. B}
  {\bf 1}, 1078(Feb 1970).

\bibitem{haviland1989}
D.~B. Haviland, Y.~Liu and A.~M. Goldman, {\em Phys. Rev. Lett.} {\bf 62}, p.
  2180(May 1989).

\bibitem{valles1989}
J.~M. Valles, R.~C. Dynes and J.~P. Garno, {\em Phys. Rev. B} {\bf 40}, p.
  6680(Octtober 1989).

\bibitem{valles1992}
J.~M. Valles, R.~C. Dynes and J.~P. Garno, {\em Phys. Rev. Lett.} {\bf 69}, p.
  3567(December 1992).

\bibitem{hebard1990}
A.~F. Hebard and M.~A. Paalanen, {\em Phys. Rev. Lett.} {\bf 65}, 927(Aug
  1990).

\bibitem{shahar1992}
D.~Shahar and Z.~Ovadyahu, {\em Phys. Rev. B} {\bf 46}, 10917(Nov 1992).

\bibitem{chervenak1999}
J.~A. Chervenak and J.~M. Valles, {\em Phys. Rev. B} {\bf 59}, p. 11209(May
  1999).

\bibitem{steiner2005}
M.~A. Steiner, G.~Boebinger and A.~Kapitulnik, {\em Phys. Rev. Lett.} {\bf 94},
  p. 107008(Mar 2005).

\bibitem{stewart2007}
M.~D. Stewart, A.~Yin, J.~M. Xu and J.~M. Valles, {\em Science} {\bf 318}, p.
  1273(nov 2007).

\bibitem{nguyen2009}
H.~Q. Nguyen, S.~M. Hollen, M.~D. Stewart, J.~Shainline, A.~Yin, J.~M. Xu and
  J.~M. Valles, {\em Phys. Rev. Lett.} {\bf 103}, p. 157001(Octtober 2009).

\bibitem{lee2011}
Y.~Lee, C.~Clement, J.~Hellerstedt, J.~Kinney, L.~Kinnischtzke, X.~Leng, S.~D.
  Snyder and A.~M. Goldman, {\em Phys. Rev. Lett.} {\bf 106}, p. 136809(Apr
  2011).

\bibitem{lin2011}
Y.-H. Lin and A.~M. Goldman, {\em Phys. Rev. Lett.} {\bf 106}, p. 127003(Mar
  2011).

\bibitem{bollinger2011}
A.~T. Bollinger, G.~Dubuis, J.~Yoon, D.~Pavuna, J.~Misewich and I.~Bozovic,
  {\em Nature} {\bf 472}, 458(April 2011).

\bibitem{sachdev_qpt}
S.~Sachdev, {\em Quantum Phase Transitions} (Cambridge, London, 1999).

\bibitem{ghosal1998}
A.~Ghosal, M.~Randeria and N.~Trivedi, {\em Phys. Rev. Lett.} {\bf 81}, p.
  3940(November 1998).

\bibitem{ghosal2000}
A.~Ghosal, M.~Randeria and N.~Trivedi, {\em Phys. Rev. B} {\bf 63}, p.
  020505(December 2000).

\bibitem{bouadim2011}
K.~Bouadim, Y.~L. Loh, M.~Randeria and N.~Trivedi, {\em Nature Physics} {\bf 7}
  (2011).

\bibitem{PhysRevLett.42.673}
E.~Abrahams, P.~W. Anderson, D.~C. Licciardello and T.~V. Ramakrishnan, {\em
  Phys. Rev. Lett.} {\bf 42}, 673(Mar 1979).

\bibitem{Lee-TVR}
P.~Lee and T.~V. Ramakrishnan, {\em Rev. Mod. Phys.} {\bf 57}, p. 287 (1985).

\bibitem{mackinnon1981}
A.~{MacKinnon} and B.~Kramer, {\em Phys. Rev. Lett.} {\bf 47}, p. 1546(November
  1981).

\bibitem{feigelman2007}
M.~V. Feigel'man, L.~B. Ioffe, V.~E. Kravtsov and E.~A. Yuzbashyan, {\em Phys.
  Rev. Lett.} {\bf 98}, p. 027001(Jan 2007).

\bibitem{fractalannals}
M.~Feigel'man, L.~Ioffe, V.~Kravtsov and E.~Cuevas, {\em Annals of Physics}
  {\bf 365}, p. 1368 (2010).

\bibitem{mirlin}
I.~Burmistrov, I.~Gornyi and A.~Mirlin, {\em Phys. Rev. Lett.} {\bf 108}, p.
  017002 (2012).

\bibitem{finkelstein1994}
A.~M. Finkel'stein, {\em Physica B} {\bf 197}, 636 (1994).

\bibitem{sacepe2011}
B.~Sac{\'e}p{\'e}, T.~Dubouchet, C.~Chapelier, M.~Sanquer, M.~Ovadia,
  D.~Shahar, M.~Feigel'man and L.~Ioffe, {\em Nat. Phys.} {\bf 7}, 239(March
  2011).

\bibitem{mondal2011}
M.~Mondal, A.~Kamlapure, M.~Chand, G.~Saraswat, S.~Kumar, J.~Jesudasan,
  L.~Benfatto, V.~Tripathi and P.~Raychaudhuri, {\em Phys. Rev. Lett.} {\bf
  106}, p. 047001(Jan 2011).

\end{thebibliography}


\end{document}